\documentclass[showpacs,aps,pra, twocolumn,amsmath,amssymb]{revtex4}

\usepackage{graphicx}
\usepackage{dcolumn}
\usepackage{bm}
\usepackage{amsmath,epsfig}

\newcommand{\ket}[1]{\left\vert#1\right\rangle}
\newcommand{\s}{\uparrow}
\newcommand{\g}{\downarrow}

\begin{document}

\author{Francesco Ciccarello\mbox{$^{1,2}$}}
\email {ciccarello@difter.unipa.it}

\author{G. Massimo Palma \mbox{$^{2}$}, Michelangelo Zarcone \mbox{$^{1}$}}
\affiliation{ \mbox{$^{1}$} CNISM and Dipartimento di Fisica e
Tecnologie Relative dell'Universit\`{a} degli Studi di Palermo,
Viale delle Scienze,
Edificio 18, I-90128 Palermo, Italy \\
\mbox{${\ }^{2}$}NEST-CNR (INFM) and Dipartimento di Scienze Fisiche ed
Astronomiche dell'Universit\`{a} degli Studi di Palermo, Via
Archirafi 36, I-90123 Palermo, Italy}

\begin{abstract} We investigate the Aharonov-Bohm (AB) interference
pattern in the electron transmission through a mesoscopic ring in
which two identical non-interacting magnetic impurities are
embedded. Adopting a quantum waveguide theory, we derive the exact
transmission probability amplitudes and study the influence of
maximally entangled states of the impurity spins on the electron
transmittivity interference pattern. For suitable electron wave
vectors, we show that the amplitude of AB oscillations in the
absence of impurities is in fact not reduced within a wide range of
the electron-impurity coupling constant when the maximally entangled
singlet state is prepared. Such state is thus able to inhibit the
usual electron decoherence due to scattering by magnetic impurities.
We also show how this maximally entangled state of the impurity
spins can be generated via electron scattering.
\end{abstract}


\title{Entanglement-induced electron coherence in a mesoscopic ring \\with two magnetic impurities}

\maketitle

\section{Introduction}

Phase coherence of electron motion is an important issue in
mesoscopic physics. Indeed, preserving coherence of the conduction
electrons is an essential requirement for the correct working of
mesoscopic semiconductor devices relying on quantum mechanical
phenomena \cite{datta,hakenb}. A well-known example of such systems
is the Aharonov-Bohm (AB) ring \cite{washburn}. Its features are due
to the the phase difference acquired by the electrons passing on the
upper and lower arms of the ring in the presence of a magnetic
field. This gives rise to the well known interference pattern in the
electron transmission as a function of the magnetic flux
\cite{washburn, xia}. Scattering by magnetic impurities, and more in
general by systems with an internal spin degree of freedom, is a
well-known source of electron decoherence
\cite{datta,imrybook,imry,schulman}. This follows from the
uncertainty in the phase shift acquired by the scattered electron
\cite{imry}. Equivalently, decoherence can be viewed as due to the
unavoidable scattering induced entanglement between the electron and
impurity degrees of freedom \cite{imry, schulman}. The transmission
properties of an AB ring with a single magnetic impurity inserted in
one of the arms and with the electron and impurity spins interacting
via a contact exchange coupling has been analyzed in \cite{joshi,
shi}. When no spin-flip occurs -- i.e. when the electron and
impurity spins are initially aligned -- the case of a static
impurity is recovered \cite{mao} and the amplitude of AB
oscillations with no impurity is preserved in a wide range of values
of the exchange coupling constant centered around zero \cite{joshi}.
However when spin-flip occurs, e.g. when the two spins are initially
anti-aligned, the amplitude of AB oscillations is reduced. This is a
signature of a loss of electron coherence which is larger for
increasing strengths of the electron-impurity coupling constant
\cite{joshi}. In this paper we consider an AB ring with two
identical spin-1/2 magnetic impurities,  one embedded in each arm.
Such system is not a mere academic extension of an AB ring with a
single impurity but shows new phenomena. A new feature that appears
in the presence of two spins is multiple scattering between the two
impurities with simultaneous occurrence of spin-flip events. This
leads to new cooperative phenomena. For instance, in the case of a
1D wire we have found that, under suitable resonance conditions,
perfect transparency appears  when the the two impurity spins are
initially prepared in the singlet maximally entangled state
\cite{ciccarello}. In the present paper we will focus on the effects
of  entanglement between the impurity spins on the AB oscillations.
Our main result is that, for suitable electron wave vectors, the
amplitude of AB oscillations turns out to be in fact not reduced in
a wide range of values of the coupling constant when the impurity
spins are prepared in the maximally entangled singlet state
$\ket{\Psi^-}$. This coherent transmission is due to an effective
suppression of spin-flip events which occurs regardless of the
electron spin state, at variance with the no spin-flip case which
occurs when the electron and impurity spins are aligned
\cite{joshi}.

This paper is organized as follows. In Sec. \ref{secapproach} we
introduce our system and illustrate the approach used to derive the
exact transmission probability amplitudes. In Sec. \ref{AB} we
investigate the AB oscillations for different spin states showing
how a nearly coherent behavior is exhibited for the singlet state of
the impurity spins. In Sec. \ref{generation} we propose a scheme for
generating this maximally entangled state via electron scattering.
The detailed mathematical derivation of all the transmission
probability amplitudes is provided in Appendix \ref{appendix}.

\section{System and Approach} \label{secapproach}

\begin{figure}[htbp]
{\hbox{{\includegraphics[scale=0.3]{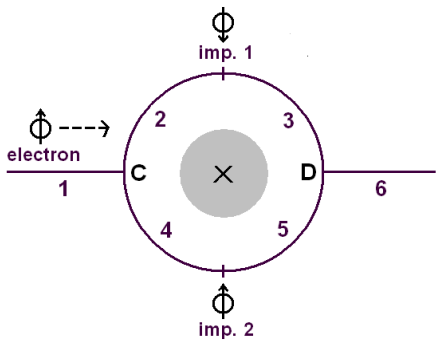}}}}
\caption{\label{ring}\footnotesize {Mesoscopic ring with two
magnetic impurities, labeled 1 and 2, inserted in the upper and
lower arm, respectively. The left and right junctions are denoted
with C and D, respectively. A static magnetic field is applied
perpendicularly to the ring plane within the gray shaded region.}}
\end{figure}

As illustrated in Fig. \ref{ring}, our system consists of a
conducting ring with equal arms. The circumference length is denoted
by $L$. We assume the width of the structure to be narrow enough to
let only the lowest electron subband to be accounted for. The
effective mass of the electron is denoted by $m^*$. Two identical
spin-1/2 impurities, labeled 1 and 2, are centered in the upper and
lower arm, respectively (see Fig. 1). The left and right junctions
of the ring are denoted by $C$ and $D$, respectively. A static
magnetic field is applied perpendicularly to the ring plane across
the internal region, excluding the wire and thus the two impurities
(see gray shaded region in Fig. 1).
As in Refs. \cite{xia, joshi, shi, mao, amjphys}, we adopt a quantum
waveguide theory approach here properly generalized to take into
account the presence of the two magnetic impurities. The local
coordinate along the electron-current direction \cite{xia} on the
upper (lower) arm is denoted as $x_1$ ($x_2$), with $0\leq x_i \leq
L/2$ ($i=1,2$). Both the above coordinates point to the junction $D$
with their origins being taken at $C$.

The Hamiltonian of the system can be written as
\begin{equation}\label{H}
H=H_0+\sum_{i=1,2}H_{ei}
\end{equation}
where $H_0$ is the Hamiltonian in the absence of impurities and
$H_{ei}$ ($i=1,2$) describes the coupling between the electron spin
$\mbox{\boldmath$\sigma$}$ and the spin-1/2 of the $i$-th impurity
$\mathbf{S}_i$. Denoting by $\mathbf{A}=\nabla \times \mathbf{B}$
the vector potential of the magnetic field $\mathbf{B}$ and by
$\mathbf{p}=-i \hbar \nabla$ the electron momentum operator, $H_0$
has the well-known form
\begin{equation}\label{H0}
H_0=\frac{1}{2m^*}\,\left(\mathbf{p}+\frac{e}{c}\,\mathbf{A}\right)^2
\end{equation}
As the vector potential is along the ring direction and has
magnitude $A=\phi/L$, with $\phi$ standing for the magnetic flux
through the ring section area, the effective representation $H_{0i}$
of (\ref{H0}) in the $i$-th arm is explicitly written as
\begin{equation}\label{schrodinger-arm1}
H_{0i}=\frac{1}{2m^*}\,\left(\frac{\hbar}{i}\frac{d}{dx_i}
+\xi\frac{e}{c}\,\frac{\phi}{L}\right)^2
\end{equation}
with $\xi=-1$ ($+1$) for $i=1$ ($2$).

We model the electron-impurity spin-spin coupling as a contact
exchange interaction according to
\begin{equation}\label{Hei}
H_{ei}=-J \, \mbox{\boldmath$\sigma$}\cdot \mathbf{S}_{i}
\,\delta(x_{i}-L/4)
\end{equation}
where $J$ is the coupling constant, $x_{i}=L/4$ is the coordinate of
the $i$-th impurity and where all the spin operators are in units of
$\hbar$. At each electron-impurity scattering event no energy
exchange takes place. However, the spin state of the two systems
will, in general, change. In particular, if the electron and
impurity spins are initially anti-aligned spin-flip may occur.

It is useful to rewrite the electron-impurity coupling Hamiltonian (\ref{Hei}) in the form
\begin{equation}\label{Hei2}
H_{ei}=-\frac{J}{2} \left(\mathbf{S}_{ei}^{2}-\frac{3}{2}\right) \delta(x_{i}-L/4)\,\,\,\,\,\,\,\,\,(i=1,2)
\end{equation}
where $\mathbf{S}_{ei}=\mbox{\boldmath$\sigma$}+ \mathbf{S}_{i}$ is
the total spin of the electron and the $i-$th impurity. Denoting by
$\mathbf{S}=\mbox{\boldmath$\sigma$}+ \mathbf{S}_{1}+\mathbf{S}_{2}$
the total spin operator, it turns out that $\mathbf{S}^2$ and $S_z$,
with quantum numbers $s$ and $m_s=-s,...,s$, respectively, are
constants of motion. It is thus convenient to use as spin space
basis the states $\ket{s_{e2}; s, m}$, common eigenstates of
$\mathbf{S}_{e2}^{2}$ (quantum number $s_{e2}$), $\mathbf{S}^{2}$
and $S_z$ \cite{nota1} (from now on, the subscript $s$ in $m_s$ will
be omitted). We denote by $t_{s_{e2}}^{(s'_{e2},s)}$ the
transmission probability amplitude that an electron injected from
the left lead with wave vector $k$ and initial spin state
$\ket{s'_{e2}; s, m}$ is transmitted in the right lead in the spin
state $\ket{s_{e2}; s, m}$. Note that, due to the form (\ref{Hei2})
of $H_{ei}$, the amplitudes $t_{s_{e2}}^{(s'_{e2},s)}$ do not depend
on $m$, as it will be made clearer in Appendix \ref{AppendixA}. The
amplitudes $t_{s_{e2}}^{(s'_{e2},s)}$ can be exactly calculated by
deriving the stationary states of the electron-two impurities
system. Due to the conservation of $\mathbf{S}^2$ and $S_z$, their
calculation can be carried on separately in each subspace of fixed
$s$ and $m$. Note that, since $\mathbf{S}_{e1}^{2}$ (quantum number
$s_{e1}$) and $\mathbf{S}_{e2}^{2}$ do not commute,
$\mathbf{S}_{e2}^{2}$ is generally not conserved. Since we are
coupling three spins 1/2, it turns out that the possible values of
$s$ are 1/2 and 3/2. When $s=1/2$ the possible values of  $s_{e2}$
are  $s_{e2}=0,1$, while for $s=3/2$ we have $s_{e2}=1$ ($s_{e1}$
can assume the same values). For left-incoming electrons with wave
vector $k$ there are eight stationary states and each of them can be
expressed as an 8-dimensional column denoted as
$\ket{\Psi_{s'_{e2};s,m}}$, where $s'_{e2}$ ($s=1/2 \Rightarrow
s'_{e2}=0,1$; $s=3/2\Rightarrow s'_{e2}=1$) is a labeling index
which generally differs from $s_{e2}$. Note that since
$s_{e1}=s_{e2}=1$ for $s=3/2$, in the subspace $s=3/2$,
$\mathbf{S}_{e1}^{2}$ and $\mathbf{S}_{e2}^{2}$ effectively commute
and thus no spin-flip can occur: the impurities behave as being
static. This is not true for the subspace $s=1/2$, for which
$s_{e1}$ and $s_{e2}$ can take values $0,1$. Therefore spin-flip in
general takes place in the subspace $s=1/2$.

The knowledge of all coefficients $t_{s_{e2}}^{(s'_{e2};s)}$, whose
detailed calculation is carried on in Appendix \ref{appendix},
completely describes the transmission properties of our system. Here
we are mainly interested in calculating how an electron with a given
wave vector $k$ and for some initial electron-impurities spin state
$\ket{\chi}$ is transmitted through the device. Thus assuming to
have the incident wave $\ket{k}\ket{\chi}$, with $\ket{\chi}$ being
an arbitrary spin state, it is straightforward to see that
$\ket{k}\ket{\chi}$ is the incoming part of the stationary state
\begin{equation}\label{sviluppochi}
\ket{\Psi_{k,\chi}}=\sum_{s_{e2}',s,m} \langle s_{e2}';s,m \ket{\chi} \ket{\Psi_{s_{e2}';s,m}}
\end{equation}
It follows that the transmitted part $\ket{\Psi_{k,\chi}}_t$ of
(\ref{sviluppochi}) provides the transmitted state into which
$\ket{k}\ket{\chi}$ evolves after its passage through the ring. To
calculate $\ket{\Psi_{k,\chi}}_t$ we simply replace each stationary
state $\ket{\Psi_{s_{e2}';s,m}}$ in (\ref{sviluppochi}) with its
transmitted part, express the latter in terms of amplitudes
$t^{(s_{e2}'; s)}_{s_{e2}}$ and rearrange (\ref{sviluppochi}) as a
linear expansion in the basis $\ket{s_{e2};s,m}$. This yields
\cite{ciccarello}
\begin{equation}\label{sviluppochi_t}
\ket{\Psi_{k,\chi}}_t=\ket{k}  \sum_{s_{e2},s,m}
\tau_{s_{e2},s,m}(\chi) \ket{s_{e2},s,m} \\
\end{equation}
with
\begin{equation} \label{alpha_coeff}
\tau_{s_{e2},s,m}(\chi) =\sum_{s_{e2}'} t^{(s_{e2}'; s)}_{s_{e2}}
\langle s_{e2}';s,m \ket{\chi}
\end{equation}
Coefficients (\ref{alpha_coeff}) fully describe how an incoming wave $\ket{k}\ket{\chi}$ is transmitted through the mesoscopic
ring. For instance, the total electron transmission coefficient $T$ can be calculated as
\begin{equation} \label{T}
T =\sum_{s_{e2},s,m} |\tau_{s_{e2},s,m}(\chi)|^2
\end{equation}

In Appendix \ref{appendix}, we derive all the coefficients $t_{s_{e2}}^{(s'_{e2},s)}$ through the calculation of the stationary
states $\ket{\Psi_{s'_{e2};s,m}}$.

\section{Aharonov-Bohm Oscillations} \label{AB}

The coefficients $t_{s_{e2}}^{(s'_{e2},s)}$ turn out to depend on
the dimensionless parameters $kL$,  $(2m^{*}/\hbar^2)(J/k)$ and
$\phi/\phi_{0}$ where   $\phi_{0}=hc/e$ is the flux quantum. In Fig.
\ref{ring} we plot the AB oscillations of electron transmission $T$
for $J/k=1,2,5$ (in unity of $2m^{*}/\hbar^2$) and $kL=1$ for an
incoming electron in the state $\ket{\s}$ and  the impurity spins in
the initial product states $\ket{\s\s}$ (a) and $\ket{\s\g}$ (b). In
the first case, the amplitude of AB oscillations is negligibly
reduced by an increase of $J/k$ in line with the no-spin flip
scattering case for one magnetic impurity \cite{joshi}. Indeed, when
the state $\ket{\s\s}$ is prepared no spin-flip takes place since
all the electron and impurity spins are aligned and the overall
state $\ket{\s}\ket{\s\s}$ belongs to the subspace $s=3/2$ where
impurities behave as being static (see Sec. \ref{secapproach}). The
behaviour of a ring with two static impurities \cite{mao} is thus
recovered. On the contrary, in the case $\ket{\s\g}$ (Fig. 2b) the
amplitude of AB oscillations is rapidly reduced for increasing
values of $J/k$ similarly to the spin-flip scattering case of Ref.
\cite{joshi}. Indeed, the initial overall spin state
$\ket{\s}\ket{\s\g}$ - which has non vanishing projection both on
subspace $s=3/2$ and on the $s=1/2$ one - is generally changed by
electron-impurities scattering into a linear combination of
$\ket{\s}\ket{\s\g}$, $\ket{\s}\ket{\g\s}$ and $\ket{\g}\ket{\s\s}$
indicating occurrence of spin-flip events. This induces loss of
coherence as witnessed by the reduction of AB oscillations with
$J/k$.
\begin{figure}[htbp]
              {\hbox{ {\includegraphics[scale=0.5]{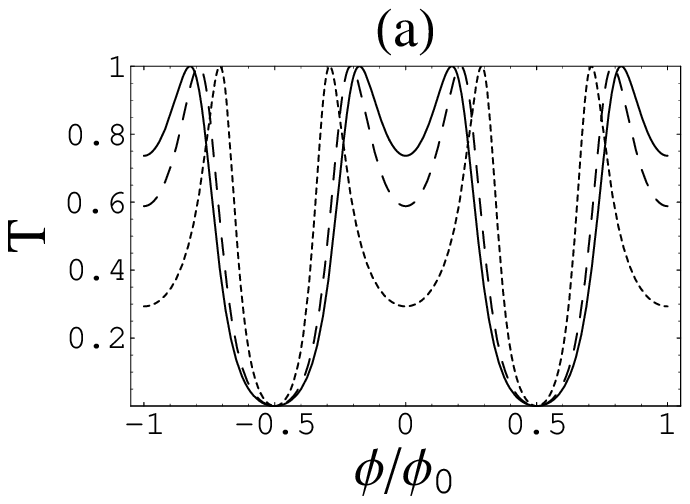}}
{\includegraphics[scale=0.5]{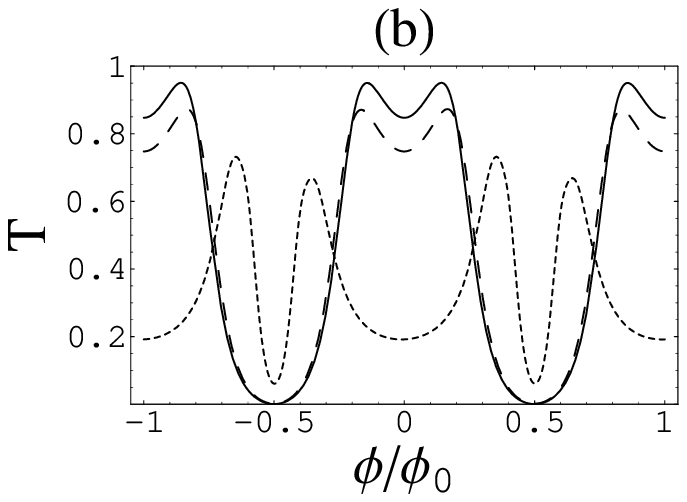}
              }}}
\caption{\label{Fig2} \footnotesize {Total transmission coefficient
$T$ versus magnetic flux $\phi/\phi_0$ for different values of $J/k$
and $kL=1$ for an incoming electron in the state $\ket{\s}$ with the
two impurities initially in the state $\ket{\s\s}$ (a) and
$\ket{\s\g}$ (b). Solid, dashed and dotted lines stand for
$J/k=1,2,5$, respectively.}} \label{fig}
\end{figure}
Denoting by $F$ the probability that the initial spin state of the
electron-impurities system is transmitted unchanged \cite{notaF}, we
define the coefficient $\eta=1-(T-F)$ such that $0\leq\eta\leq 1$.
Since for $T=F$ $\eta=1$ it turns out that $\eta$ provides
information of occurrence of spin-flip or not: high values of $\eta$
correspond to low probability of spin-flip events. For instance, in
the case of Fig. 2a $\eta\simeq 1$ for any value of $\phi$ and $J/k$
since spin-flip does not take place at all. In Fig. \ref{eta}a
$\eta$ is plotted as a function of $\phi/\phi_0$ for $J/k=1,2,5$ and
for the initial state $\ket{\s}\ket{\s\g}$. As $J/k$ is increased,
the minimum value of $\eta$ gets lower and lower due to higher
chance for spin-flip to occur.

We now investigate electron transmission when the impurity spins are
initially in a maximally entangled state. Here we focus on the
maximally entangled triplet and singlet states
$\ket{\Psi^{\pm}}=(\ket{\s\g}\pm \ket{\g\s})/\sqrt{2}$. Considering
an incoming electron in the state $\ket{\s}$, Fig. \ref{Fig3} shows
the behaviour of $T$ for $\ket{\Psi^{+}}$ (a) and $\ket{\Psi^{-}}$
(b).
\begin{figure}[htbp]
              {\hbox{ {\includegraphics[scale=0.5]{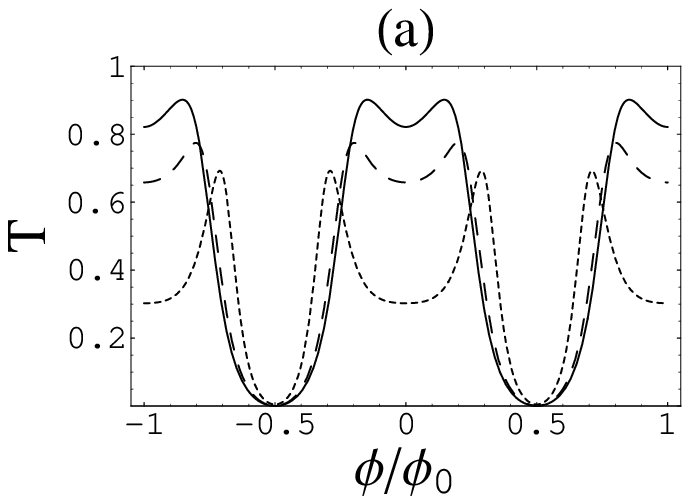}}
{\includegraphics[scale=0.5]{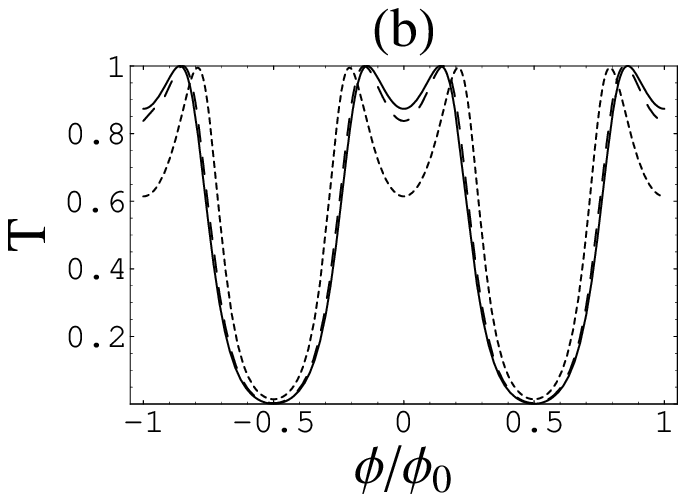}
              }}}
\caption{\footnotesize {Total transmission coefficient $T$ versus
magnetic flux $\phi/\phi_0$ for different values of $J/k$ and $kL=1$
for an incoming electron in the state $\ket{\s}$ with the two
impurities initially in the state $\ket{\Psi^+}$ (a) and
$\ket{\Psi^-}$ (b). Solid, dashed and dotted lines stand for
$J/k=1,2,5$, respectively.}}
\end{figure}
In the case of $\ket{\Psi^+}$ a behaviour similar to Fig. 2b with
reduction of the amplitude of AB oscillations, signature of the
presence of decoherence, appears. This is confirmed by an analysis
of $\eta$ under the same conditions (Fig. \ref{eta}b) showing
occurrence of spin-flip.

However, a striking behaviour is observed for the singlet state
$\ket{\Psi^-}$. Despite the fact that the spin state
$\ket{\s}\ket{\Psi^-}$ fully lies in the subspace $s=1/2$ (where
spin-flip may occur) the oscillations' amplitude is negligibly
reduced for the considered strengths of $J/k$, resembling
qualitatively the coherent behaviour of Fig. 2a. Indeed, as showed
in Fig. \ref{eta}c, $\eta\simeq 1$ in this case. An effective
suppression of spin-flip and decoherence thus takes place provided
the impurity spins are in the state $\ket{\Psi^-}$. Unlike the no
spin-flip case of Fig. \ref{Fig2}a requiring alignment of the
electron and impurity spins, our calculations show that, as it will
be made clearer later, the present behavior is exhibited regardless
of the spin state of the incoming electron. No constraint on the
spin polarization of the incident electron is thus required for this
phenomenon to occur.
\begin{figure}[htbp]
{\hbox{ {\includegraphics[scale=0.35]{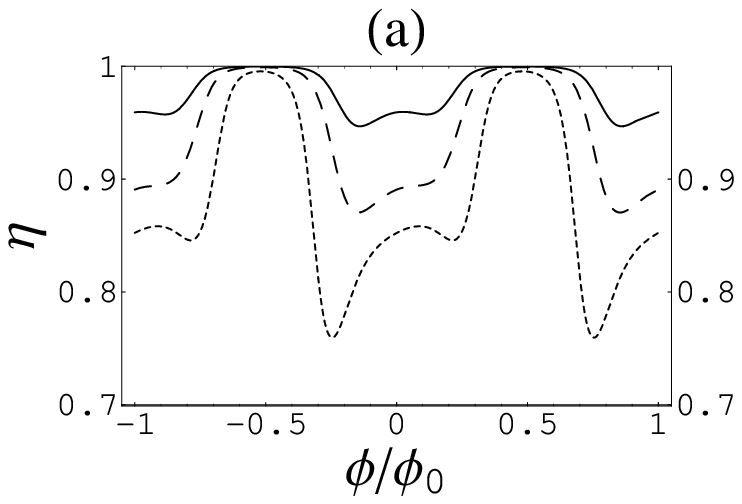}}
{\includegraphics[scale=0.35]{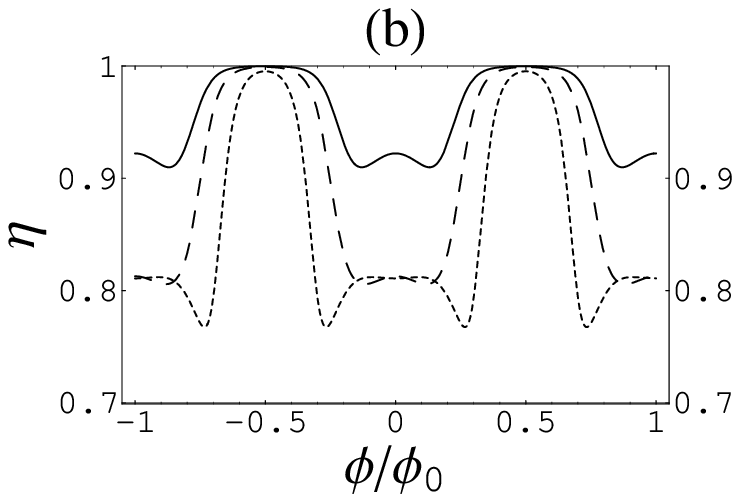}
              {\includegraphics[scale=0.35]{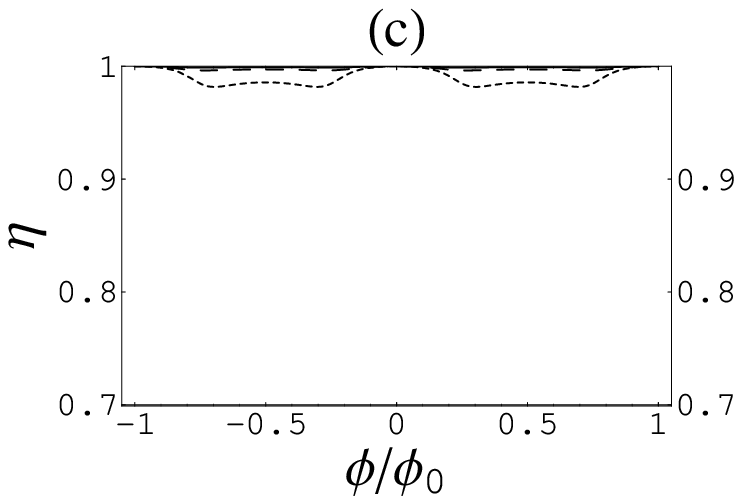}} }}}
\caption{\label{eta} \footnotesize {$\eta=1-(T-F)$ vs. magnetic flux
$\phi/\phi_0$ for the initial spin state $\ket{\s}\ket{\s\g}$ (a),
$\ket{\s}\ket{\Psi^+}$ (b) and $\ket{\s}\ket{\Psi^-}$ (c). Solid,
dashed and dotted lines stand for $J/k=1,2,5$, respectively. The
$\eta$ axis starts from 0.7.}}
\end{figure}
The amplitude of AB oscillations for a fixed strength of $J/k$ can
be calculated as the difference between the maximum and minimum of
$T$ over the interval $-1\leq \phi/\phi_0\leq 1$. In Fig. 5 we plot
this amplitude as a function of $J/k$ for $kL=1$ and for the initial
spin states $\ket{\s}\ket{\s\s}$, $\ket{\s}\ket{\s\g}$,
$\ket{\s}\ket{\Psi^+}$ and $\ket{\s}\ket{\Psi^-}$. As pointed out
previously, in the case of $\ket{\s}\ket{\s\s}$ the problem reduces
to a potential scattering with two static impurities and there
exists a finite range of values of $J/k$ centered at $J/k=0$ where
the amplitude is not reduced \cite{joshi, mao}. This is not the case
for $\ket{\s}\ket{\s\g}$ and $\ket{\s}\ket{\Psi^+}$ for which the
amplitude never equals 1 even at small $J/k$ and gets smaller and
smaller for increasing strengths of $J/k$. Similarly to the no-spin
flip case $\ket{\s}\ket{\s\s}$, for $\ket{\s}\ket{\Psi^-}$ there is
a finite range around $J/k=0$ showing negligible amplitude
reduction.
\begin{figure}[htbp]
{\hbox{ {\includegraphics[scale=0.3]{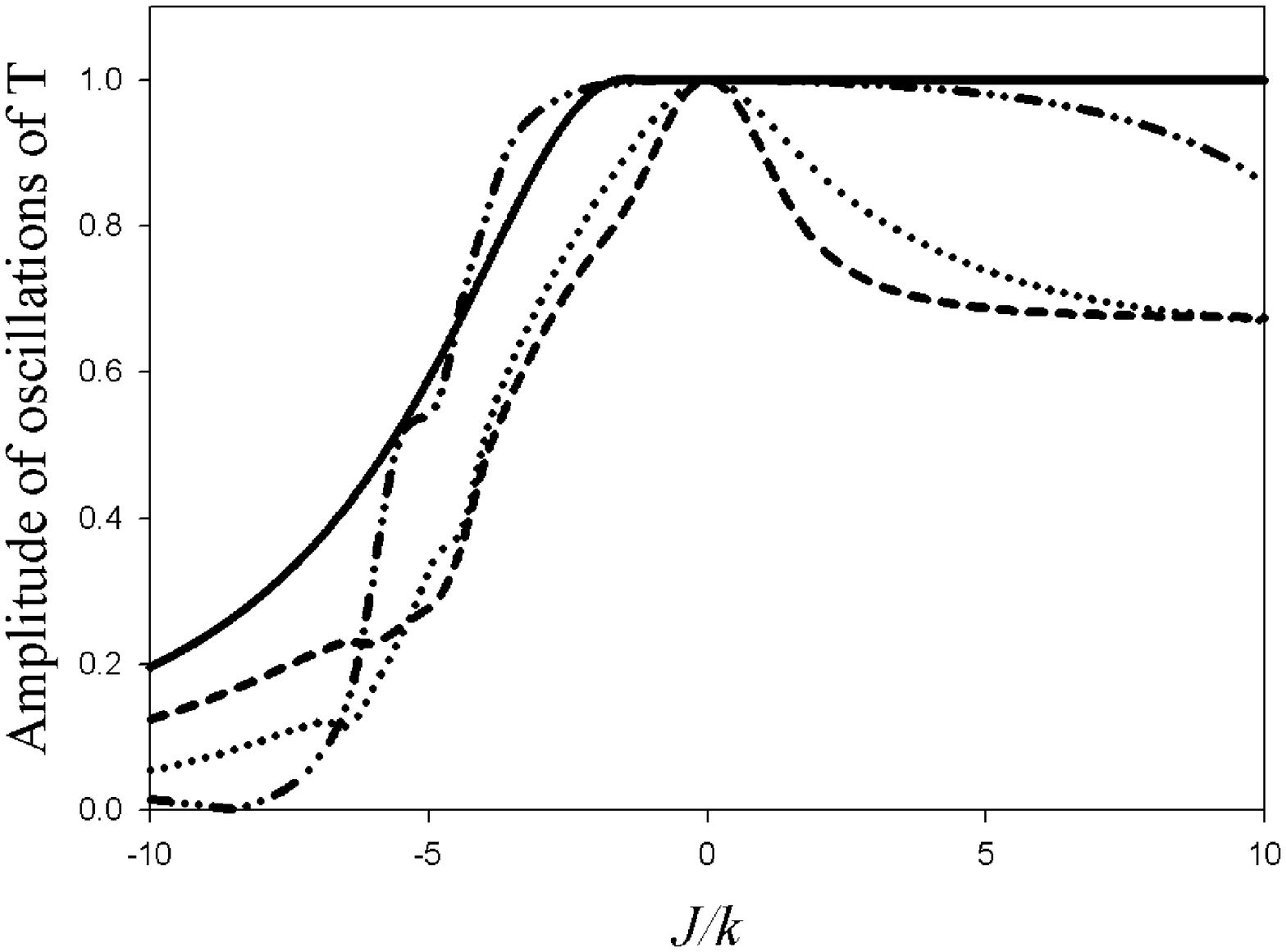}}}} \caption{\footnotesize
{\label{Fig5} Amplitude of Aharonov-Bohm oscillations vs. $J/k$ for
$kL=1$ and in the case of the initial spin states
$\ket{\s}\ket{\s\s}$ (------), $\ket{\s}\ket{\s\g}$ ($\cdot \cdot
\cdot \cdot \cdot$), $\ket{\s}\ket{\Psi^+}$ (--- ---) and
$\ket{\s}\ket{\Psi^-}$ ($- \cdot \cdot \,-$).}}
\end{figure}

\begin{figure}
 \includegraphics [scale=0.4]{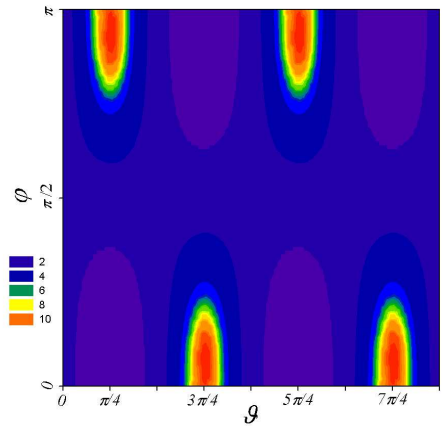}
 \caption{\label{teta_phi} (Color online) $R$ as a function of $\theta$ and $\varphi$ at $kL=1$
 when the electron is injected in the up spin state $\ket{\s}$
 with the impurities prepared in the state
$\cos{\theta}\ket{\s\g}+e^{i\varphi}\sin{\theta}\ket{\g\s}$.}
\label{fig4}
\end{figure}

To quantify the above finite range, we introduce the quantity
$R$, defined as the width of the interval around $J/K =
0$ where the amplitude of AB oscillation for a given initial spin
state is larger than $0.95$. To further illustrate how the electron transmission through the
ring depends on the state in which the two impurities are prepared
let us consider the family of one spin up impurity spins states
\begin{equation} \label{family}
\ket{\Psi (\theta , \varphi)} = \cos\theta\ket{\s\g} +
e^{i\varphi}\sin\theta\ket{\g\s}
\end{equation}
with $\theta\in[0,2\pi]$ and $\varphi\in[0,\pi]$. This family
includes both maximally entangled and product states. In Fig.
\ref{teta_phi} we plot $R$ as a function of $\theta$ and $\varphi$
when the electron is injected in the spin state $\ket{\s}$ and the
impurities are prepared in a state (\ref{family}) for $kL= 1$.
Note how the width of the range showing a nearly coherent behaviour
depends crucially on the relative phase $\varphi$ between the
impurity spin states $\ket{\s\g}$ and $\ket{\g\s}$. Four sharp
maxima of $R$ appear in correspondence to the impurity spins
prepared in the singlet state $\ket{\Psi^{-}}$. These results
suggests that $\varphi$ can be regarded as a sort of control
parameter of electron coherence in a mesoscopic ring with two spin
1/2 impurities.

\begin{figure}[htbp]
{\hbox{{\includegraphics[scale=0.4]{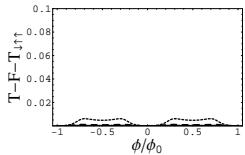}}}}
\caption{\label{Fig7}\footnotesize {$T-F-T_{\s\g\g}$ vs.
$\phi/\phi_0$ for $kL=1$ and the initial spin state
$\ket{\s}\ket{\Psi^+}$. Solid, dashed and dotted lines stand for
$J/k=1,2,5$, respectively.}}
\end{figure}

\begin{figure}
 \includegraphics [scale=0.3]{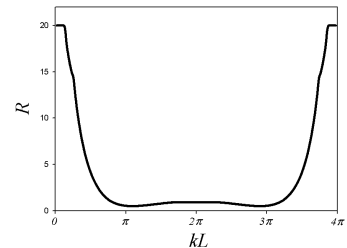}
 \caption{\label{R_kl} $R$ vs. $kL$ in the case of the initial spin
state $\ket{\s}\ket{\Psi^-}$ for $kL\in[0,4\pi]$.}
\end{figure}

Such behavior can be explained in terms of an effective
quasi-conservation of $\mathbf{S}_{12}^2$, where
$\mathbf{S}_{12}=\mathbf{S}_{1}+\mathbf{S}_{2}$ denotes the total
spin of the two impurities. This observable have quantum number
$s_{12}=0,1$, corresponding to the singlet and triplet subspaces,
respectively. According to such quasi-conservation, whose origin we
will explain shortly, and taking into account the fact that $S_{z}$
is a constant of motion, the initial spin state
$\ket{\s}\ket{\Psi^+}$ is scattered into a linear combination of
$\ket{\s}\ket{\Psi^+}$ and $\ket{\g}\ket{\s\s}$. However, since the
eigenspace $s_{12}=0$, $m=1/2$ is non degenerate, the initial spin
state $\ket{\s}\ket{\Psi^-}$ is transmitted or reflected nearly
unchanged, as shown Fig. \ref{eta}c.  The same of course is true for
$\ket{\g}\ket{\Psi^-}$ and thus for any state
$\left(\alpha\ket{\s}+\beta\ket{\g}\right)\ket{\Psi^-}$ with
arbitrary complex values of $\alpha$ and $\beta$ (this explains why
no constraint on the spin polarization of the incoming electron is
required). The validity of this quasi-conservation law is further
confirmed by Fig. \ref{Fig7} where, for  the initial state
$\ket{\s}\ket{\Psi^+}$, we plot the difference between $T$ and the
sum of $F$ and $T_{\g\s\s}$, with $T_{\g\s\s}$ being the probability
that the system is transmitted in the state $\ket{\g}\ket{\s\s}$. A
maximum value not larger than 1$\%$ is reached.
The origin of quasi-conservation of $\mathbf{S}_{12}^2$ can be
explained as follows. The wavefunction in the upper arm of the ring
is a linear combination of $e^{ik_{1}x}$ and $e^{-ik_{2}x}$, while
in the lower one it is linear combination of $e^{ik_{2}x}$ and
$e^{-ik_{1}x}$ with $k_1=k+(e\phi/\hbar c L)$ and
$k_2=k-(e\phi/\hbar c L)$ (see Appendix \ref{AppendixA}). This
induces an asymmetry between the two arms when a flux $\phi$ is
present, since in this case $k_1$ differs from $k_2$. When $\phi=0$
perfect symmetry holds and $\mathbf{S}_{12}^2$ must be rigorously
conserved. Indeed, for $\phi=0$ $\ket{\s}\ket{\Psi^-}$ is
transmitted perfectly unchanged (in Fig. \ref{eta}c $\eta=1$ for
$\phi=0$), while $\ket{\s}\ket{\Psi^+}$ can only either remain
unchanged or be flipped into $\ket{\g}\ket{\s\s}$ (in Fig.
\ref{Fig7} $T-F-T_{\s\g\g}$ exactly vanishes for $\phi=0$). In the
presence of a flux $\phi$, the above symmetry is broken and perfect
conservation of $\mathbf{S}_{12}^2$ generally does not occur.
However, even for $\phi\neq 0$ as $kL$ approaches $4n\pi$
($n=0,1,...$) the symmetry still holds. Indeed, the boundary
conditions at the ring junctions and those at the two impurities'
positions imply that the stationary states of the system depend on
$kL$ through phase factors of the form $e^{\pm ikL/4}$ and $e^{\pm
ikL/2}$ (see Appendix \ref{AppendixA}). It is thus straightforward
to see that, for each stationary state, the squared modulus of the
wave function is a periodic function of $kL$ with period $4\pi$,
this meaning that for $kL=4n\pi$ with $n\neq 0$ the transmission
properties of the ring coincide with those obtained for $kL=0$. In
the regime $kL=4n\pi$ the system thus behaves as if
$k_1=-k_2=e\phi/\hbar c L$ and the two arms turn out to be
symmetric. To illustrate this, in Fig. \ref{R_kl} we plot $R$ versus
$kL$ the range $[0,4\pi]$ in the case of the initial spin state
$\ket{\s}\ket{\Psi^-}$. The above mentioned range around $J/k=0$
where a nearly vanishing amplitude reduction takes place shows an
increasing width as $kL$ approaches $0$ or $4\pi$ as a signature of
an increasing symmetry of the two arms of the ring. For
discrepancies between $kL$ and $4n\pi$ up to $0.8\pi$ the width of
the coherent range is still appreciable (in Fig. \ref{R_kl} $R > 1$
for $kL < 0.8\pi$). Good robustness against deviation from the
symmetry condition $kL=4n\pi$ is thus exhibited.
This also explains why the effect of survival of the AB
oscillations' amplitude for the singlet state of the impurity spins
is observable in Figs. 3b and \ref{Fig5} -- where we have considered
the representative value $kL=1$ -- since it turns out that
$kL=1\simeq 0.3\, \pi$.

\section{Generation of Maximally Entangled States} \label{generation}

In this section we mainly address the issue of how to generate the
maximally entangled state $\ket{\Psi^-}$ of the two impurity spins
giving rise to the above described effects. Possible implementations
of a magnetic impurity in the single-channel ring are a paramagnetic
impurity atom having a virtual state in the continuum \cite{amjphys}
or a quantum dot with one excess unpaired electron and thus behaving
as an effective spin 1/2 system \cite{joshi,ssqc}.

Here we propose a method for entangling the two impurities via electron scattering with an entanglement mediator. Generation of
entangled states of two magnetic impurities via electron scattering in 1-dimensional systems has been recently investigated in
\cite{ciccarello, yang, yasser, giorgi, nakazato}.

\begin{figure}
 \includegraphics [scale=0.6]{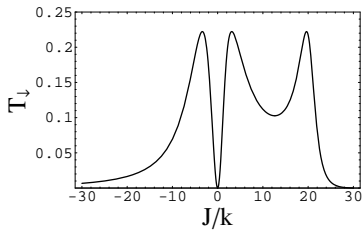}
 \caption{\label{tdown} Spin down transmission coefficient $T_{\g}$ at $kL=1$ and $\phi=0$ as a function of
 $J/k$ when the electron is injected in the state $\ket{\s}$ with the impurities
prepared in the state $\ket{\g\g}$.}
\end{figure}

The basic idea of these schemes is to send an electron in the up
spin state $\ket{\s}$ with the two impurity spins initially in a
product state $\ket{\g\g}$. The impurities are so far apart that
their mutual coupling is negligible. The electron interacts with
each scatterer via an exchange interaction. $S_z$ is conserved in
the scattering process and the transmitted spin state will result in
a linear combination of $\ket{\s}\ket{\g\g}$, $\ket{\g}\ket{\Psi^+}$
and $\ket{\g}\ket{\Psi^-}$. If the transmitted electron is filtered
in the down spin state $\ket{\g}$, the two impurities are generally
left in an entangled state. This state is not necessarily maximally
entangled \cite{yasser}. However, under conditions allowing
$\mathbf{S}_{12}^2$ to be an additional constant of motion the above
scheme always projects the impurities in the maximally entangled
state $\ket{\Psi^+}$ \cite{ciccarello}. Of course, once
$\ket{\Psi^+}$ has been generated, $\ket{\Psi^-}$ can be easily
obtained by simply introducing a relative phase shift through a
local field acting on one of the two impurities. For our system
(Fig. \ref{ring}), this method works when $\phi=0$ since, as
discussed in Sec. \ref{AB}, in this regime both $S_{z}$ and
$\mathbf{S}_{12}^2$ are strictly conserved. Denoting by $T_{\g}$ the
spin polarized probability that the electron is transmitted in the
state $\ket{\g}$ (that is to project the impurities in the state
$\ket{\Psi^+})$, in Fig. \ref{tdown} we plot $T_{\g}$ versus $J/k$
in the case $kL=1$.
A probability higher than 20$\%$ can be reached with $|J/k|$ lower
than 5. In particular, a maximum is exhibited at $J/k\simeq 3$ that
is well within the range where the amplitude of AB oscillations for
the singlet state is in fact not reduced (see Fig. \ref{Fig5}). We
have checked that similar behaviors yielding analogous maximum
probabilities to generate $\ket{\Psi^+}$ take place for different
values of $kL$.

We point out that while in the case of a 1-dimensional wire the
above scheme works correctly only for electron wave vectors allowing
conservation of $\mathbf{S}_{12}^2$ \cite{ciccarello}, in the
present case such constraint is not required. This is due to the
previously discussed symmetry of the ring in the absence of an
applied flux (see Sec. \ref{AB}).

\section{Conclusions}

In this work, we have considered a mesoscopic ring with two
identical spin 1/2 magnetic impurities embedded one in each arm at
symmetric locations. Electrons entering the wire undergo multiple
scattering by the impurities giving rise to spin-flip processes
before being definitively reflected or transmitted. Developing a
proper quantum waveguide theory approach based on the coupling of
three angular momenta, the exact stationary states and thus all the
transmission probability amplitudes of the system in the presence of
a magnetic flux have been derived. In agreement with previous
studies, we have shown that occurrence of spin-flip events resulting
from electron-impurities scattering generally reduces the amplitude
of AB oscillations, as a signature of the well-known magnetic
impurities induced-electron decoherence. However, an anomalous
behaviour appears in the case of the singlet maximally entangled
state $\ket{\Psi^-}$ of the impurity spins. At suitable incident
electron wave vectors, the amplitude of AB oscillations turns out to
be negligibly reduced in a finite range of the electron-impurities
coupling constant. At the same time spin-flip turns out to be nearly
frozen. This survival of electron coherence via entanglement of the
localized spins occurs regardless of the spin state of the incoming
electron and thus no constraint on the electron spin polarization is
required. Such behaviours have been explained in terms of a
quasi-conservation law of the squared total spin of the two
impurities. We have proposed a scheme for generating the maximally
entangled states $\ket{\Psi^{\pm}}$ of the two impurities through
the same electron scattering mechanism.

In line with the case of a 1-dimensional wire where a perfect
resonance condition is always found for two impurities in the state
$\ket{\Psi^-}$ \cite{ciccarello}, we have thus shown how this
maximally entangled state is able to effectively ``freeze" the usual
decoherence due to scattering with magnetic impurities. This
phenomenon strongly reminds the so called \emph{decoherence free
subspaces} (DFS), a well-known topic in the study of open quantum
systems \cite{PalmaSE96}, but usually dealt with for systems
encountered in quantum optics. Our work can thus be regarded as a
manifestation of DFS in one of the most familiar mesoscopic devices.

\appendix

\section{\label{AppendixA} \textbf{Derivation of the transmission probability amplitudes}}
\label{appendix}

In this Appendix we derive all the transmission probability
amplitudes $t_{s_{e2}}^{(s'_{e2},s)}$ required for calculating the
transmission properties of the system for a given initial spin state
$\ket{\chi}$ according to (\ref{sviluppochi_t}) and
(\ref{alpha_coeff}). This requires the calculation of the stationary
states $\ket{\Psi_{s'_{e2};s,m}}$. We basically adopt a quantum
waveguide theory approach \cite{xia,mao} here properly generalized
to take into account the presence of two magnetic impurities. We
first consider the subspace $s=3/2$ and then the subspace $s=1/2$.

\subsection{Subspace $s=3/2$}

In this 4-dimensional subspace ($m=-3/2,...,3/2$), both $s_{e1}$ and
$s_{e2}$ can have the only possible value 1. It follows that in this
case $\mathbf{S}_{e1}^{2}$ and $\mathbf{S}_{e2}^{2}$ effectively
commute and the four stationary states $\ket{\Psi_{1; 3/2,m}}$
belonging to this subspace are thus eigenstates of
$\mathbf{S}_{e2}^{2}$ taking the form
\begin{equation}
\ket{\Psi_{1;3/2,m}}=\ket{\varphi}\ket{1; 3/2, m}
\end{equation}
where $\ket{\varphi}$ belongs to the electron orbital space. In this
case $s_{e2}$ is a good quantum number and coincides with $s'_{e2}$.
Moreover, the effective form of $H_{ei}$ reads
\begin{equation} \label{Hei_23/2}
H_{ei}=-J/4 \, \delta(x_{i}-L/4)
\end{equation}

Eq. (\ref{Hei_23/2}) shows that in this subspace the two impurities
behave as if they were static and thus no spin-flip may occur. The
standard case of a ring with two static impurities of Ref.
\cite{mao} is thus recovered. The wave function $\varphi_{i}(x)$ in
each segment $i=1,...,6$ (see Fig. \ref{ring}) \cite{nota2} can be
easily derived by solving the Schr\"{o}dinger equation
$H_0\ket{\varphi}=E\ket{\varphi}$ and obtaining \cite{xia}
\begin{subequations}
\begin{eqnarray}
\varphi_{1}(x)&=&\alpha_{1}e^{ikx}+\beta_{1}e^{-ikx}  \label{wavefunction1}\\
\varphi_{m}(x)&=&\alpha_{m}e^{ik_{1}x}+\beta_{m}e^{-ik_{2}x}\,\,\,\,\,(m=2,3)\\
\varphi_{n}(x)&=&\alpha_{n}e^{ik_{2}x}+\beta_{n}e^{-ik_{1}x}\,\,\,\,\,\,\,\,(n=4,5)\\
\varphi_{6}(x)&=& t_{1}^{(1,3/2)}e^{ikx}  \label{wavefunction4}
\end{eqnarray}
\end{subequations}
with $k=\sqrt{2m^*E}/\hbar$ and
\begin{eqnarray}\label{k1_k2}
k_{1}=k+(e\phi/\hbar c L) \\
k_{2}=k-(e\phi/\hbar c L)
\end{eqnarray}
Setting $\alpha_{1}$ to unity, the other unknown coefficients in
Eqs. (\ref{wavefunction1})-(\ref{wavefunction4}) $\alpha$, $\beta$
and the transmission probability amplitude $t_{1}^{(1,3/2)}$ can be
determined by imposing proper boundary conditions on the wave
function and its derivative at junctions $C$ and $D$ and at the two
impurities' sites \cite{xia, joshi, mao}.

Matching of the wave function at these points as well as
conservation of current density at $C$ and $D$ must hold. This
yields the following boundary conditions

\begin{subequations}
\begin{eqnarray}
\varphi_{1}(0)&=&\varphi_{2}(0) \label{boundcond1_1}\\
\varphi_{1}(0)&=&\varphi_{4}(0)\\
\varphi_{3}(L/2)&=&\varphi_{6}(0)\\
\varphi_{5}(L/2)&=&\varphi_{6}(0)\\
\varphi_{2}(L/4)&=&\varphi_{3}(L/4)\\
\varphi_{4}(L/4)&=&\varphi_{5}(L/4)\\
\varphi'_{1}(0)&=&\varphi'_{2}(0)+\varphi'_{4}(0)\\
\varphi'_{6}(0) &=& \varphi'_{3}(L/2)+\varphi'_{5}(L/2)
\label{boundcond1_8}
\end{eqnarray}
\end{subequations}
On the other hand, at the two impurities' positions the derivative
of the wave function shows a discontinuity due to the $\delta$-like
potential (\ref{Hei_23/2}) according to conditions
\begin{subequations}
\begin{eqnarray}
\varphi'_{2}(L/4)-\varphi'_{3}(L/4)&=&\frac{2m^*}{\hbar^2} \frac{J}{4}\,\varphi_{2}(L/4)  \label{boundcond2_1}\\
\varphi'_{4}(L/4)-\varphi'_{5}(L/4)&=&\frac{2m^*}{\hbar^2}\frac{J}{4}\,\varphi_{4}(L/4)
\label{boundcond2_2}
\end{eqnarray}
\end{subequations}
which can be easily derived by integrating the Schr\"{o}dinger
equation across the two impurities' locations \cite{mao}.

Boundary conditions (\ref{boundcond1_1})-(\ref{boundcond1_8}) and
(\ref{boundcond2_1})-(\ref{boundcond2_2}) form a linear system in
the unknown coefficients appearing in
(\ref{wavefunction1})-(\ref{wavefunction4}). Once this is solved,
the transmission amplitude $t_{1}^{(1,3/2)}$ can be obtained.
$t_{1}^{(1,3/2)}$ (whose lengthy expression is not shown here) does
not depend on $m$ due to the effective form (\ref{Hei_23/2}) of
$H_{ei}$.

\subsection{Subspace $s=1/2$}

In this 4-dimensional subspace $s_{e1},s_{e2}=0,1$ and thus
$\mathbf{S}_{e1}^{2}$ and $\mathbf{S}_{e2}^{2}$ do not commute and
$s_{e2}$ is not a good quantum number. This is a signature of the
fact that in this space spin-flip may occur. Therefore, for each
fixed value of $m=-1/2,1/2$ the stationary states are not
eigenstates of $\mathbf{S}_{e2}^{2}$ and take the form
\begin{equation}  \label{ansatz}
\ket{\Psi_{s'_{e2};1/2,m}}=\sum_{s_{e2}=0,1}\ket{\varphi_{s'_{e2},s_{e2}}}\ket{s_{e2};
1/2, m}
\end{equation}
where $\ket{\varphi_{s'_{e2},s_{e2}}}$ are orbital wave functions.
Note that in the present subspace $s'_{e2}\neq s_{e2}$. For fixed
$s'_{e2}=0,1$ $\ket{\Psi_{s'_{e2};1/2,m}}$ can be found by solving
the Schr\"{o}dinger equation. It is straightforward to see that in
each segment $i=1,...,6$  the two wave functions
$\varphi_{s'_{e2},0}(x)$ and $\varphi_{s'_{e2},1}(x)$ turn out to
take a form analogous to Eqs.
(\ref{wavefunction1})-(\ref{wavefunction4}). For each of them,
continuity of the wave functions as well as conservation of current
density at junctions $C$ and $D$ must hold similarly to the case
$s=3/2$. However, in this case appropriate boundary conditions on
the derivatives of the wave functions at the impurities'sites have
to be derived. To do this, we consider the Schr\"{o}dinger equation
at the ring arm containing the $i-$th impurity. This reads
\begin{widetext}
\begin{eqnarray}\label{scrhrodinger}
\left[\frac{\hbar^2}{2m^{*}}\frac{d^2}{dx_{i}^2}-\frac{1}{2m^*}\left(\frac{\xi
e \phi}{cL}\right)^2+i\frac{\hbar}{m^*}\frac{\xi
e\phi}{cL}\frac{d}{dx_{i}} +\frac{J}{2} \left(\mathbf{S}_{ei}^{2}  -
\frac{3}{2}\right)
\delta(x_{i}-L/4)+E\right] \Psi_{s'_{e2};1/2,m}(x_i)=0\nonumber\\
\end{eqnarray}

Integration of both sides of Eq. (\ref{scrhrodinger}) across the
impurity position yields
\begin{equation}\label{scrhrodingerint}
\left[\Delta
\Psi'_{s'_{e2};1/2,m}(L/4)+\frac{2m^{*}}{\hbar^2}\frac{J}{2}
\left(\mathbf{S}_{ei}^{2}-\frac{3}{2}\right)
\Psi_{s'_{e2};1/2,m}(L/4)\right]=0
\end{equation}
\end{widetext}
where $\Delta \Psi'_{s'_{e2};1/2,m}(L/4)$ stands for the jump of the
derivative at the impurity's site. Once expansion (\ref{ansatz}) is
inserted into Eq. (\ref{scrhrodingerint}) and this is projected onto
$\ket{0;1/2,m}$ and $\ket{1;1/2,m}$, we obtain the following
boundary conditions

\begin{subequations}
\begin{eqnarray}
\Delta \varphi_{s_{e2}',0}'(L/4)&=&- \frac{2m^*J}{\hbar^{2}} \frac{\sqrt{3}}{4}\,\varphi_{s_{e2}',1}(L/4)\label{boundcondacc1}\\
\Delta
\varphi_{s_{e2}',1}'(L/4)&=&\frac{2m^*J}{\hbar^{2}}\frac{1}{2}\,\varphi_{s_{e2}',1}(L/4)\nonumber\\
&-&  \frac{2m^*J}{\hbar^{2}}\frac{\sqrt{3}}{4}
\,\varphi_{s_{e2}',0}(L/4)\label{boundcondacc2}
\end{eqnarray}
\end{subequations}
at impurity 1 and

\begin{subequations}
\begin{eqnarray}
\Delta \varphi_{s_{e2}',0}'(L/4)&=&\frac{2m^*J}{\hbar^{2}}\frac{3}{4}\,\varphi_{s_{e2}',0}(L/4)  \label{boundcondacc3}\\
\Delta\varphi_{s_{e2}',1}'(L/4)&=&-\frac{2m^*J}{\hbar^{2}}\,\frac{1}{4}\,\varphi_{s_{e2}',1}(L/4)\label{boundcondacc4}
\end{eqnarray}
\end{subequations}

at impurity 2. In deriving Eqs. (\ref{boundcondacc1})
-(\ref{boundcondacc4}) we have used that

\begin{subequations}
\label{matrixel}
\begin{eqnarray}
\langle 0;1/2,m \left|S_{e1}^2\right|0;1/2,m
\rangle&=&\frac{3}{2}\\
\langle 0;1/2,m \left|S_{e1}^2\right|1;1/2,m \rangle&=&\frac{\sqrt{3}}{2}\\
\langle 1;1/2,m \left|S_{e1}^2\right|0;1/2,m \rangle&=&\frac{\sqrt{3}}{2}\\
\langle 1;1/2,m \left|S_{e1}^2\right|1;1/2,m \rangle&=&\frac{1}{2}
\end{eqnarray}
\end{subequations}
The above matrix elements of $\mathbf{S}_{ei}^{2}$ can
be easily computed by means of $6j$-coefficients, these allowing to
go from the scheme $e2$, where the electron is first coupled to
impurity 2, to the $e1$ one.

Note how $\varphi_{s_{e2}',0}(x)$ and $\varphi_{s_{e2}',1}(x)$ turn
out to be mutually coupled by boundary conditions
(\ref{boundcondacc1}) -(\ref{boundcondacc4}).
Finally, also taking into account the above mentioned boundary
conditions analogous to Eqs.
(\ref{boundcond1_1})-(\ref{boundcond1_8}), one ends up with a linear
system in 20 unknown variables. This can be solved for each
$s'_{e2}=0,1$ to obtain the transmission amplitudes
$t_{0}^{(s'_{e2},1/2)}$ and $t_{1}^{(s'_{e2},1/2)}$.

Note that these do not depend on $m$ due to the form (\ref{Hei2}) of
$H_{ei}$ and to the fact that $6j$ coefficients and thus matrix
elements (\ref{matrixel}) are $m$-independent (see for instance
\cite{weissbluth}).

It is important to point out that our calculations yield
$t^{(s_{e2}'; 1/2)}_{s_{e2}}\neq 0$ for $s_{e2}\neq s_{e2}'$, as a
signature of non conservation of $\mathbf{S}_{e2}^{2}$ and
definitively of occurrence of spin-flip in the subspace $s=1/2$. The
analytical formulas obtained for $t_{0}^{(s'_{e2},1/2)}$ and
$t_{1}^{(s'_{e2},1/2)}$ are quite lengthy and will not be shown
here.

\begin{acknowledgements}
Helpful discussions with Y. Omar (Technical University of Lisbon),
V. R. Vieira (Instituto Superior T\'{e}cnico, Lisbon) and G. Falci
(Universit\`{a} degli Studi di Catania) are gratefully acknowledged.
\end{acknowledgements}

\end{document}